\documentclass[%
 reprint,
 amsmath,amssymb,
 aps,
prl,
]{revtex4-1}
	\usepackage{feynmf}
	\usepackage{graphics}
	\usepackage{epsfig}
	\usepackage{graphicx}
	\usepackage{color}
	\usepackage{comment}

\begin{document}

\title{Importance of Spin-Orbit Coupling in Organic BEDT-TTF and BEDT-TSF Salts}

\author{Stephen M. Winter}
 \email{winter@physik.uni-frankfurt.de}
\author{Kira Riedl}
\author{Roser Valent\'i}
\affiliation{Institut fur Theoretische Physik, Goethe-Universitat Frankfurt, 60438 Frankfurt am Main, Germany}

\begin{abstract}
We investigate the spin-orbit coupling (SOC) effects in $\alpha$- and
$\kappa$-phase BEDT-TTF and BEDT-TSF organic salts. 
Contrary to the assumption  that SOC  in organics is negligible due to
 light C, S, H atoms, we show the relevance of such an interaction in
a few representative cases. In the weakly
correlated regime, SOC manifests primarily in the opening of energy gaps at
degenerate band touching points. This effect becomes especially important
 for  Dirac semimetals such as $\alpha$-(ET)$_2$I$_3$. 
Furthermore, in the magnetic insulating phase,
SOC results in additional anisotropic exchange interactions, which provide a
compelling explanation for the controversial 
field-induced behaviour of the quantum
spin-liquid candidate $\kappa$-(ET)$_2$Cu$_2$(CN)$_3$. 
  We conclude by discussing the importance of SOC for
the description of low-energy properties in organics.
\end{abstract}


\maketitle

Layered organic materials have long served as ideal systems for studying
complex physical phenomena such as the interplay between charge and spin order,
unconventional superconductivity, and, exotic phases
emerging from strong electronic correlations\cite{dressel2011quantum,powell2006strong,lebed2008physics,toyota2007low}. In these systems, high
compressibility and synthetic versatility allows fine tuning of the underlying
interactions via physical and chemical pressure, thus providing access to many
different ground states. For example, significant evidence has recently emerged
for a quantum spin liquid (QSL) state in a number of triangular-lattice
organics, including $\kappa$-(ET)$_2$Cu$_2$(CN)$_3$ and
$\kappa$-(ET)$_2$Ag$_2$(CN)$_3$\cite{test,PhysRevLett.112.177201,yamashita2008thermodynamic,yamashita2010highly,PhysRevLett.117.107203}. However, the appearance of very small
energy gaps, and field-induced anomalies in the former material remain
essentially unexplained. Likewise, interest has recently grown in the
$\alpha$-phase materials $\alpha$-(ET)$_2$I$_3$\cite{tajima2000transport} and $\alpha$-(BETS)$_2$I$_3$\cite{inokuchi1995electrical}.
The former material has been argued to form, under pressure, a zero gap
semimetal (ZGS) with the Fermi energy located at the
intersection of a pair of tilted Dirac
cones\cite{tajima2006electronic,tajima2000transport,kajita2014molecular}.
However, the low-energy electronic and magnetic response shows significant
departure from theoretical expectations for simple Dirac systems \cite{cryst2020643,hirata2016observation}. An unexplored aspect is the effect of
spin-orbit coupling (SOC) in these systems. 

For (inorganic) 
magnetic insulators, strong SOC
may generate large anisotropic magnetic interactions, associated with exotic
spin-liquid states discussed e.g. for the iridates and $\alpha$-RuCl$_3$\cite{rau2014generic,winter2016challenges,jackeli2009mott,banerjee2016proximate,PhysRevLett.108.127204,johnson2015monoclinic,chaloupka2010kitaev}. In weakly correlated systems such as BiSb, Bi$_2$Te$_3$, etc. \cite{chen2009experimental,hasan2010colloquium,RevModPhys.88.021004,RevModPhys.83.1057},
 SOC tends to open a gap in the bulk, and may lead to nontrivial band topology and associated exotic edge
states. Such effects have rarely been considered for the organic ET salts, as the light
C, S, H atoms provide only moderate SOC. Nonetheless, it is known that SOC plays a dominant role in the spin relaxation of graphene\cite{balakrishnan2013colossal,huertas2006spin,huertas2009spin} and organic-based semiconductors\cite{PhysRevLett.110.216602}, and may be relatively enhanced for orbitally degenerate systems\cite{merino2016topological}. Indeed, in this work, we argue,
despite a weak relative magnitude, that SOC is relevant for
 the
{\it low-energy}  properties in selected (representative) organic salts.
 We discuss as primary
examples, the $\alpha$-phase Dirac semimetals, and $\kappa$-phase spin liquid
materials.

At first approximation, the effect of SOC is to introduce a spin-dependent, complex hopping\cite{bernevig2013topological}:
\begin{align}
\mathcal{H}_{\text{hop}} =& \  \sum_{ij} \mathbf{c}_{i}^\dagger (t_{ij} \mathbb{I}_{2\times 2} + \frac{i}{2}\vec{\lambda}_{ij}\cdot \vec{\sigma})\mathbf{c}_{j}
\end{align}
in terms of the single particle operators $\mathbf{c}_{i}^\dagger =
\left(c_{i,\uparrow}^\dagger \ \ c_{i,\downarrow}^\dagger \right)$, where
$c^\dagger_{i,\sigma}$ creates an electron at molecular site $i$, with spin
$\sigma \in\{\uparrow,\downarrow\}$ and $t_{ij}$ describes hopping between the
highest occupied molecular orbital (HOMO) at sites $i$ and $j$. The vector
quantities $\vec{\lambda}_{ij} = -\vec{\lambda}_{ji}$ arise from SOC;
$\vec{\sigma} = (\sigma_x,\sigma_y,\sigma_z)$ is the Pauli vector and
$\mathbb{I}_{2\times 2}$ is the $2\times 2$ identity matrix. Estimates of $t_{ij}$ and
$\vec{\lambda}_{ij}$ (Supplementary table \ref{table-hops}) were obtained for various ET and BETS
salts at the density functional theory (DFT) level using
methods implemented in the ORCA code~\footnote{See supplemental materials, available at XX}. For S-based ET salts
$|\vec{\lambda}_{ij}|\sim 1-2$ meV, while SOC is stronger in the Se-based BETS
salts and $|\vec{\lambda}_{ij}|\sim 5-10$ meV. In each case,
 $\vec{\lambda}_{ij}$ tends to point along the long axis of the ET
or BETS molecules, regardless of the details of the crystal
 packing~\cite{Note2}; $\vec{\lambda}_{ij}$ is also largest for molecules $i,j$ with molecular planes
oriented 90$^\circ$ to each other, suggesting the strongest SOC effects occur
for e.g. $\alpha$- and $\kappa$-phase salts.

 We first consider the $t,|\vec{\lambda}| \gg U$ weakly correlated limit, where
SOC alters the dispersion and spin-orbital composition of the single particle
band states. In general, given $|\vec{\lambda}/t|\ll 1$ in the organics, SOC
effects are strongest near band-touching $\mathbf{k}$-points, at
which the Hamiltonian including SOC is: \begin{align}
\mathcal{H}_\mathbf{q} =  \left(\mathbf{c}^\dagger_{\mathbf{q},+} \ \ \mathbf{c}^\dagger_{\mathbf{q},-} \right) \left(\begin{array}{cc}\epsilon_\mathbf{q}\ \mathbb{I}_{2\times 2}& i\vec{\lambda}_\mathbf{q} \cdot \vec{\sigma} \\ -i\vec{\lambda}^*_\mathbf{q} \cdot \vec{\sigma} & \epsilon_\mathbf{q} \ \mathbb{I}_{2\times 2}\end{array} \right)\left(\begin{array}{c}\mathbf{c}_{\mathbf{q},+}\\ \mathbf{c}_{\mathbf{q},-}\end{array}\right)\label{eqn-2}
\end{align}
where the upper $(+)$ and lower $(-)$ bands become degenerate at $\mathbf{k} =
\mathbf{q}$ in the absence of SOC. With SOC, the bands are split into two
spin-orbital pairs, with energy difference $\Delta = 2|\vec{\lambda}_\mathbf{q}|$. Only
the spin component parallel to $\vec{\lambda}_{\mathbf{q}}$ is conserved, which
strongly modifies the magnetic response of electrons in the vicinity of such points. For example, in the
absence of SOC, an external field $\mathbf{H}$ induces an energy splitting $\Delta E =
g\mu_B|\mathbf{H}|$ between spins aligned parallel and antiparallel to the field. In the
presence of SOC, the splitting becomes $\Delta E_\mathbf{q} = (C_+ + C_-)/2$,
where $C_\pm =
\sqrt{(g\mu_Bh_{||}\pm|\vec{\lambda}_\mathbf{q}|)^2+(g\mu_Bh_\perp)^2}$ and
$h_{||}$ and $h_\perp$ are the components of $\mathbf{H}$ parallel and perpendicular to
$\vec{\lambda}_\mathbf{q}$, respectively.
 Defining $g_\text{eff} =
\lim_{h\rightarrow 0}(1/\mu_B) (\partial \Delta E / \partial h)$ yields
$g_{\mathbf{q},||} = g \sim 2$ and $g_{\mathbf{q},\perp} = 0$. That is, a
perpendicular field imparts {\it no} Zeemen splitting of the
$\mathbf{k}=\mathbf{q}$ states. For this reason, we anticipate an
anomalous angular dependence of the spin susceptibility $\chi_s$ and related
NMR Knight shift for states close to $\mathbf{k}=\mathbf{q}$. These
observations are extremely relevant, for example, to the $\alpha$-phase Dirac semimetals
$\alpha$-(ET)$_2$I$_3$ and $\alpha$-(BETS)$_2$I$_3$\cite{inokuchi1995electrical,tajima2006electronic,cryst2020643,tajima2006electronic,kajita2014molecular}. 
Unlike graphene, the Dirac points in these materials
are not symmetry protected; via Eq. (\ref{eqn-2}),
SOC is predicted to open a direct band-gap $\Delta \sim 2|\vec{\lambda}_\mathbf{q}| \sim 1-2$
meV for $\alpha$-(ET)$_2$I$_3$, and $\Delta \sim 5-10$ meV for the related $\alpha$-(BETS)$_2$I$_3$.
The presence of this direct gap implies a true Dirac ZGS state is impossible with SOC. 
Indeed, for $\alpha$-(ET)$_2$I$_3$, the existence of a small gap (or pseudo gap) $\sim 1$ meV is indicated by a suppressed conductivity $\sigma(T)$\cite{tajima2006electronic,PhysRevLett.116.226401} and spin susceptibility\cite{hirata2016observation} observed over a wide pressure range below $10-20$ K. For $\alpha$-(BETS)$_2$I$_3$ a suppression of $\sigma(T)$ is
observed below 50 K at 3 kbar~\cite{inokuchi1995electrical}, while at 6 kbar, a
constant carrier density was observed below 20 K, consistent with an indirect negative gap
semimetal~\cite{tajima2006electronic}. While these observations have been discussed in terms of residual charge-order\cite{PhysRevB.93.195116} and correlation effects\cite{PhysRevLett.116.226401,hirata2016observation}, SOC represents an equally relevant effect at the energy scale of the observed gaps, and therefore must be considered in analysis of the $\alpha$-phase semimetals.
\begin{figure}[t]
\includegraphics[width=\linewidth]{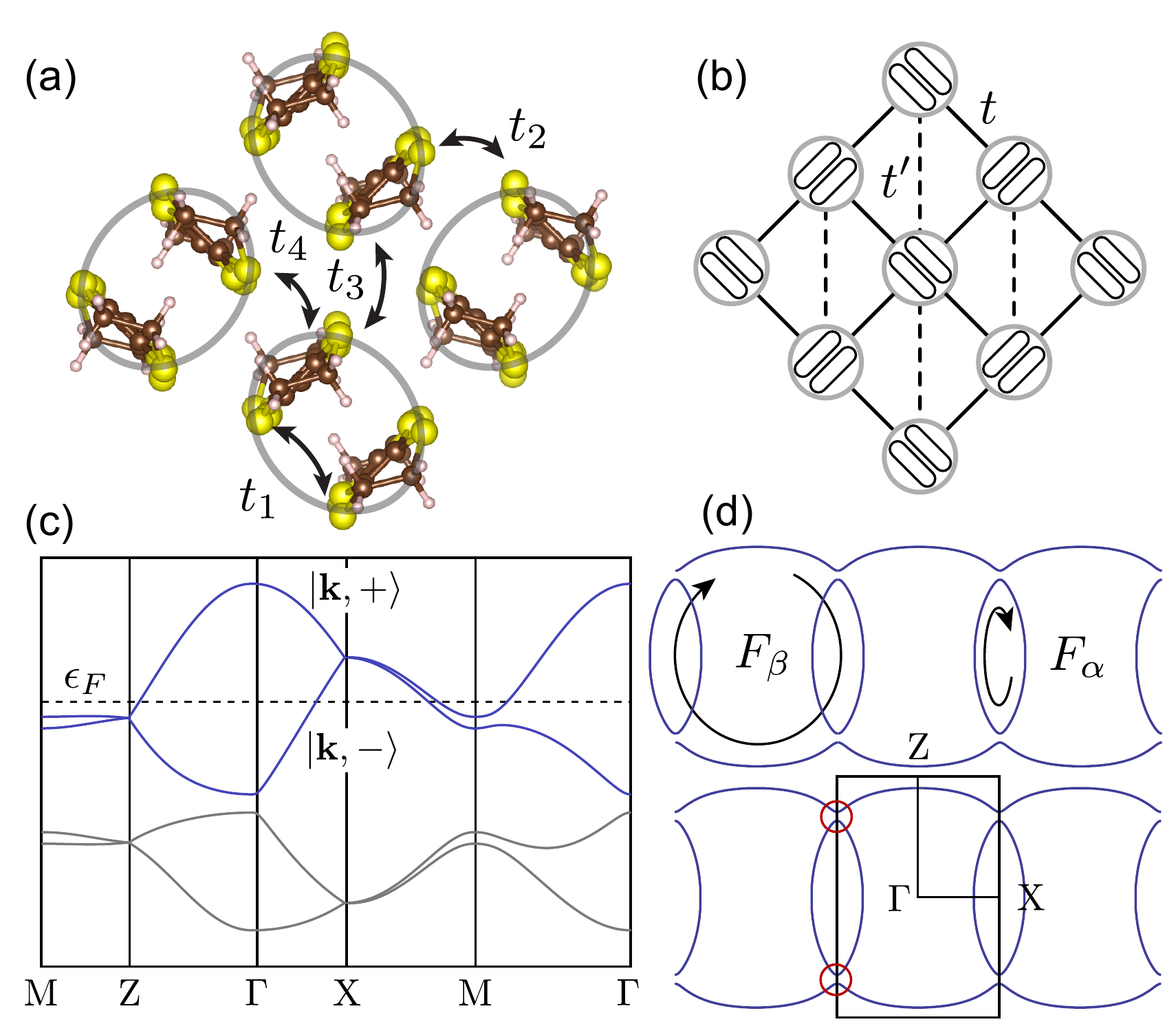}
\caption{\label{fig-magnetic} (a) Definition of molecular hopping integrals $t_{1-4}$ for $\kappa$-phase salts compared to (b) dimer-dimer hopping integrals $t,t^\prime$. Including SOC modifies the dispersion (c) and Fermi surfaces (d), opening a gap along the zone-edge X$\rightarrow$M and M$\rightarrow$Z.
In these figures we use hopping parameters for $\kappa$-(BETS)$_2$Cu[N(CN)$_2$]Br, but take $\lambda_n$ five times greater than estimated, in order to emphasize the effects.}
\end{figure}

For the $\kappa$-phase salts (Fig.~\ref{fig-magnetic} (a)),
 the Fermi level intersects two bands associated
with the antibonding combination of HOMOs at each molecular dimer (Fig.
\ref{fig-magnetic} (c)). Without SOC, these bands would be degenerate at the
Brillouin zone boundary \cite{guterding2015influence,kandpal2009revision,PhysRevB.89.045102}, but SOC splits the bands. This splitting
may be directly observed in quantum oscillation experiments. Such experiments
typically reveal two frequencies $F_\alpha$ and $F_\beta$ with associated
amplitude $I_\alpha$ and $I_\beta$ (Fig.
\ref{fig-magnetic} (d))~\cite{PhysRevB.59.12370,kartsovnik2004high}. The first
corresponds to orbit around the small hole pocket. The second represents  the
total area of the first Brillouin zone, and requires tunnelling between the
hole pockets and open sheets through the spin-orbit gaps at the zone edge. The
tunnelling rate is given by $\tau = \exp(-H_0/2H)$, where the magnetic
breakdown field $H_0$ is related to the tunnelling barrier by $H_0 \propto
\Delta_{\text{eff}}^2$. For $H\ll H_0$, $\tau \sim 0$, and only the
$\alpha$-orbit is observable ($I_\alpha \gg I_\beta$). For $H\gg H_0$, $\tau
\sim 1$, and $I_\alpha \ll I_\beta$. In the absence of SOC, but retaining
inversion symmetry, $\Delta_\text{eff} \sim 0$, and $I_\alpha$ is expected to
vanish at all fields. Nonetheless, $F_\alpha$ oscillations are observed
in $\kappa$-phase salts, often with $I_\alpha \ll I_\beta $ in ET
salts\cite{PhysRevB.59.12370,Ohmichi:1998vi,mielke1997fermi} and $I_\alpha
\gtrsim I_\beta$ in Se-based BETS
salts~\cite{Steven:2009ie,Pesotskii:1999tj,Hill:1998vb}.
Assuming that
$\Delta_\text{eff}$ arises primarily from SOC, we expect the ratio of breakdown
fields $H_0$(BETS)/$H_0$(ET) $\sim (\lambda_{Se}/\lambda_S)^2\sim 25$, where
$\lambda_S$ and $\lambda_{Se}$ are the atomic spin-orbit constants of sulfur
and selenium, respectively \cite{murov1993handbook}. We may estimate the SOC-induced gap as $\Delta_\text{eff} \sim
2|\vec{\lambda}_\mathbf{q}|\sim 1$ meV for ET salts 
corresponding to $H_0$(ET)
$\lesssim 1-2$ T, far below typical quantum oscillation fields (larger
than $10$ T). For
BETS salts, $\Delta_\text{eff} \sim 5-10$ meV suggesting $H_0$(BETS)
$\lesssim 20-50$ T, on par with that of $\kappa$-(ET)$_2$Cu(SCN)$_2$, for which
inversion symmetry breaking introduces a non-SOC gap~\cite{Goddard:2003hp}.  
Therefore, the apparent experimental observation that $H_0$(BETS) $\gg H_0$(ET) in a variety of salts is fully consistent with 
SOC playing a dominant role in the magnetic breakdown
behaviour of the $\kappa$-phase salts.

\begin{table}[t]
\caption {\label{table-Js} Computed exchange interactions $J$ and DM vectors
$\mathbf{D}$~\cite{Note2}; all values are in units of K. Values for the
$Pnma$ salts are in the coordinate system $(a,b,c)$, while the $P2_1/c$ values
are with respect to $(a,b,c^*)$ (see Fig.~\ref{fig-cant}).} \begin{ruledtabular}
\begin{tabular}{ccccc}
Material&$J$&$J^\prime$&$\mathbf{D}$\\
\hline
$\kappa$-(ET)$_2$Cu[N(CN)$_2$]Cl&482&165&(-3.65, -3.58, -0.17)\\
$\kappa$-(ET)$_2$Cu$_2$(CN)$_3$&228&268&(+3.30, +0.94, +0.99)\\
$\kappa$-(ET)$_2$Ag$_2$(CN)$_3$&250&157&(-2.93, -0.92, -2.93)\\
$\kappa$-(ET)$_2$B(CN)$_4$&131&365&(+1.03, +4.17, -0.08)\\
\end{tabular}
\end{ruledtabular}
\end{table}
We now consider the $\kappa$-phase materials in the insulating $U,t_1\gg t_{2-4}$ limit, where a single hole is localized on each dimer, occupying the antibonding combination of HOMOs. Including SOC, the Hamiltonian for the localized spins up to $\mathcal{O}(t^2)$ is:
\begin{align}
\mathcal{H} = & \ \sum_{ij} J_{ij} \ \mathbf{S}_i \cdot \mathbf{S}_j + \mathbf{D}_{ij} \cdot \mathbf{S}_i \times \mathbf{S}_j + \mathbf{S}_i \cdot \mathbf{\Gamma}_{ij} \cdot \mathbf{S}_j
\end{align}
where $J_{ij}$ is the Heisenberg coupling between dimer
sites $i,j$. SOC introduces $\mathbf{D}_{ij}$, the
Dzyaloshinskii-Moriya (DM) vector, and $\mathbf{\Gamma}_{ij}$, the pseudo-dipolar tensor\cite{moriya1960anisotropic,yildirim1995anisotropic}. The ET dimers form an anisotropic triangular lattice; by convention,
we label values for the solid bonds in Fig. \ref{fig-magnetic}(b) as $J$, $\mathbf{D}$, $\mathbf{\Gamma}$, and values for the dashed bonds as $J^\prime$,
$\mathbf{D}^\prime$, $\mathbf{\Gamma}^\prime$. Due to the presence of a
crystallographic inversion centre, $\mathbf{D}^\prime = 0$, and $\mathbf{\Gamma}^\prime \sim 0$. Estimates for $J, J^\prime$, and $\mathbf{D}$ (Table \ref{table-Js}) were
obtained for selected salts via cluster exact diagonalization~\cite{Note2}. 

\begin{figure}[t]
\includegraphics[width=\linewidth]{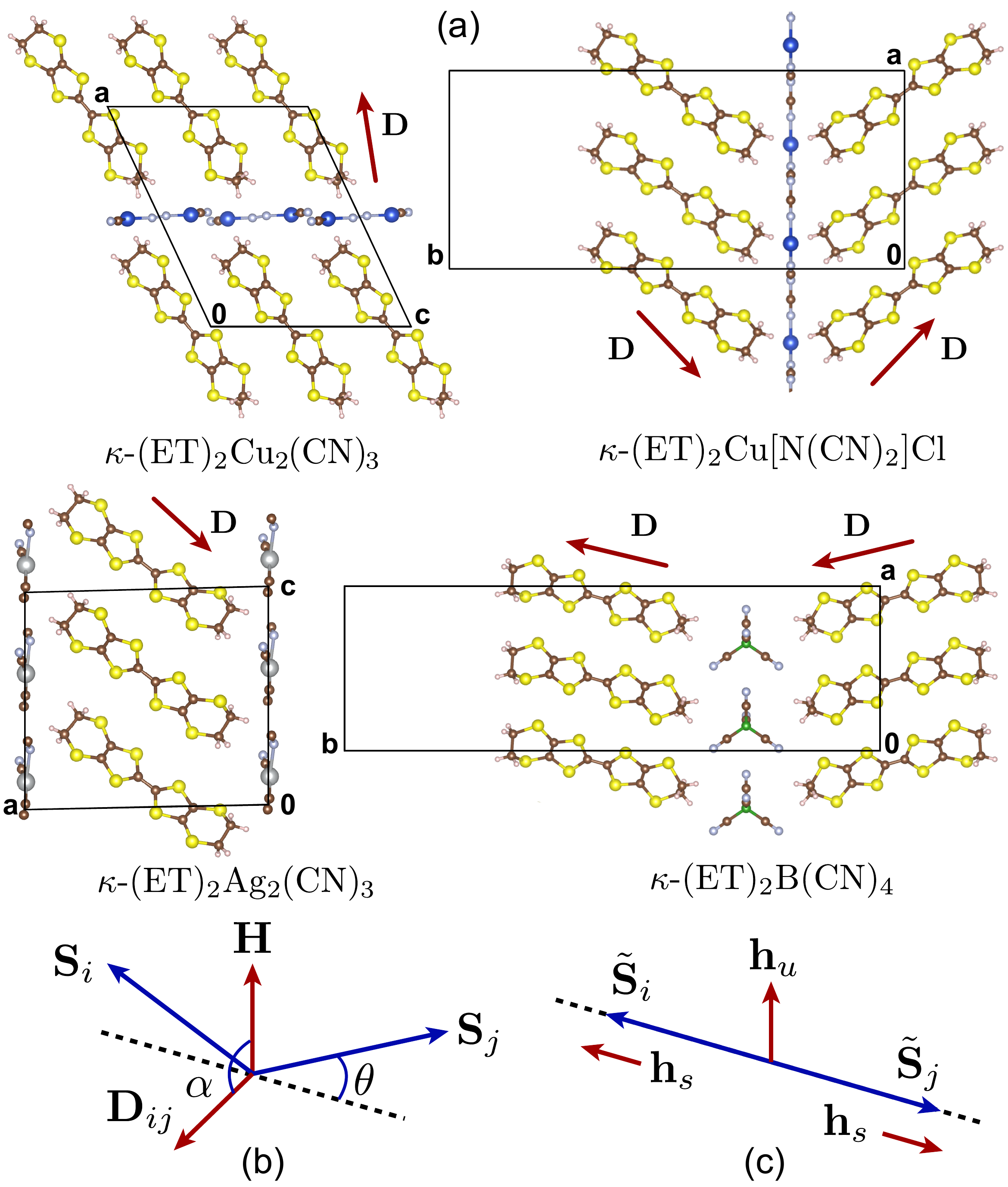}
\caption{\label{fig-cant} (a) Computed orientation of $\mathbf{D}$ in the plane perpendicular to the $2_1$ axis for selected ET salts. In each case, $\mathbf{D}$ is nearly along the long-axis of the molecule. The presence of $\mathbf{D}_{ij}$ leads to spin-canting by an angle $\theta$, as shown in (b). This canting may be ``gauged away'' by making local rotations around $\pm\theta$, generating an effective staggered field $\mathbf{h}_s$ (c).}
\end{figure}

Of the salts studied, $\kappa$-(ET)$_2$Cu[N(CN)$_2$]Cl exhibits the largest $J/J^\prime \sim 5$, consistent with observed square lattice N\'eel order below $T_N = $ 27 K\cite{miyagawa1995antiferromagnetic,lunkenheimer2012multiferroicity}. The smallest $J/J^\prime \sim 0.3$ belongs to $\kappa$-(ET)$_2$B(CN)$_4$, which exhibits quasi 1D (spin-liquid) behaviour down to $T = $ 5 K \cite{yoshida2015spin}. The 2D spin-liquid candidates $\kappa$-(ET)$_2$Cu$_2$(CN)$_3$\cite{shimizu2003spin,yamashita2008thermodynamic} and $\kappa$-(ET)$_2$Ag$_2$(CN)$_3$\cite{PhysRevLett.117.107203,pinteric2016anions} have $J/J^\prime \sim 1-2$, close to the triangular limit. The orientation of
each DM-vector with respect to the ET molecules is shown in Fig. \ref{fig-cant}(a); in each case $\mathbf{D}$ is nearly parallel to
$\vec{\lambda}_{ij}$, along the long axis of the ET molecules.

For the $\kappa$-phase salts, all
local $\mathbf{D}_{ij}$ are nearly collinear (Fig.~\ref{fig-cant} (a)), since the component along the
$2_1$ axis direction is typically small. If we initially ignore this component,
the anisotropic terms can be nearly ``gauged away'' by making local rotations
of $\mathbf{S}_i$ around $\mathbf{D}_{ij}$ by a canting angle $\theta =
\frac{1}{2}\eta\cos^{-1} (\frac{4J^2-|\mathbf{D}|^2}{4J^2+|\mathbf{D}|^2})$,
where $\eta=+1(-1)$ for different sublattices (Fig.~\ref{fig-cant} (b)-(c)). The rotated Hamiltonian is thus:
\begin{align}
\mathcal{H} =& \  \sum_{ij}\tilde{J}_{ij}\tilde{\mathbf{S}}_i\cdot\tilde{\mathbf{S}}_j + g\mu_B\sum_{i} (\mathbf{h}_u+\eta\mathbf{h}_s)\cdot\tilde{\mathbf{S}}_i
\end{align}
with modified exchange coupling $\tilde{J}_{ij} =
J_{ij}+\frac{|\mathbf{D}_{ij}|^2}{4J_{ij}}$, effective uniform external field
$\mathbf{h}_u\approx \mathbf{H}$, and a staggered field
$\mathbf{h}_s(\perp\mathbf{h}_u)$ with periodicity $\mathbf{q}=(\pi,\pi)$ that
goes as $\mathbf{h}_s = \frac{1}{2J_{ij}} \ (\mathbf{H}\times
\mathbf{D}_{ij})$.  Anisotropy in the local $g$-tensor also contributes to
$\mathbf{h}_s$ with periodicity $(\pi,\pi)$. As pointed out by many previous
works \cite{miyagawa1995antiferromagnetic,lunkenheimer2012multiferroicity,pinteric1999magnetic}, the staggered DM-interaction promotes a canted moment in the plane
perpendicular to $\mathbf{D}$ for N\'eel ($\pi,\pi$) order, as observed in
$\kappa$-(ET)$_2$Cu[N(CN)$_2$]Cl. For this case, detailed NMR and ESR
measurements below $T_N$ suggested that $\mathbf{D}$ lies close to the
$ab$-plane making an angle of $\phi_a \sim 45^\circ$ from the
$a$-axis\cite{smith2004precise,smith2003dzialoshinskii,antal2009spin}. ESR
measurements further provided a magnitude $|\mathbf{D}| \sim 5.0$ K, and $J
\sim 600$ K\cite{antal2009spin}. These values are in good agreement with the
computed values of $\phi_a = 44.5^\circ$ and $|\mathbf{D}| = 5.1$ K, and
$J=480$ K.

\begin{figure}[t]
\includegraphics[width=\linewidth]{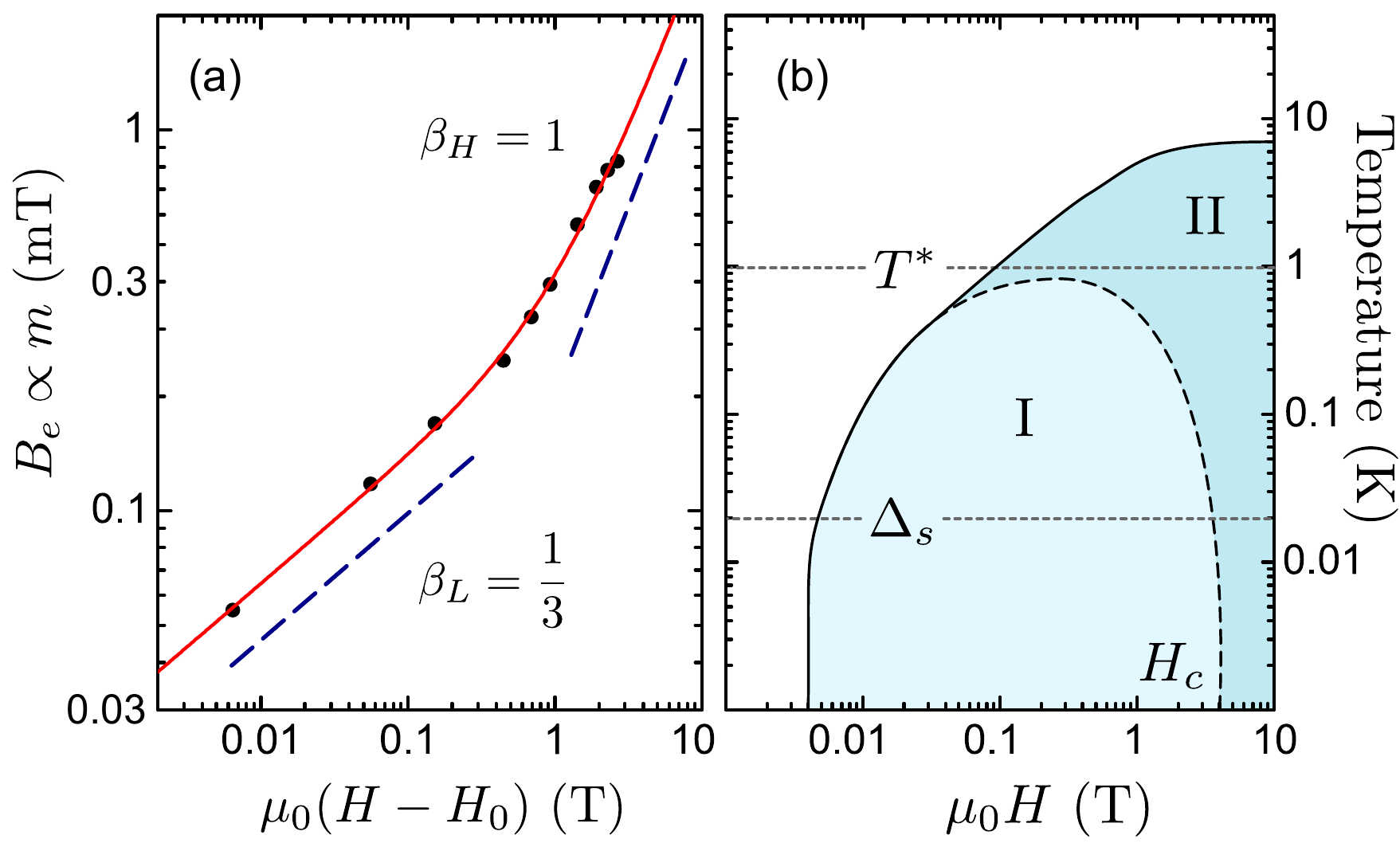} \caption{\label{fig-muon} (a) Fit
of the 0.8 K $\mu$SR linewidth of $\kappa$-(ET)$_2$Cu$_2$(CN)$_3$ from 
\cite{pratt2011magnetic} to $B_e \propto \int d\Omega \ m$, yielding
$\beta_L=\frac{1}{3}$ and $\beta_H = 1$. The slope change near 1 T corresponds
to a crossover between the staggered-field dominated region (I) at low field to
uniform-field dominated region (II) at high field, as shown in (b).}
\end{figure}

The effects of the staggered field $\mathbf{h}_s$ on the purported 2D quantum
spin-liquid (QSL) states of $\kappa$-(ET)$_2$Cu$_2$(CN)$_3$ and
$\kappa$-(ET)$_2$Ag$_2$(CN)$_3$ have not yet been considered. Intuition can be gained by recalling the effect in 1D AFM
chains\cite{oshikawa1997field,zhao2003effects,affleck1999field,oshikawa2002electron}, for which $\mathbf{h}_s$ has been studied in great detail. For the purpose of discussion, we first ignore
the components of $\mathbf{D}$ along the $2_1$ axis, and consider the effects
of a staggered field at $\mathbf{q}=(\pi,\pi)$. In 1D QSLs, the staggered
susceptibility $\chi_s \sim H^{\gamma_s}$, $\gamma_s \sim -\frac{2}{3}$ is
typically divergent at low field, and greatly exceeds the uniform
susceptibility $\chi_u\sim H^{\gamma_u}$, $\gamma_u\sim 0$. Assuming the
magnetizations follow $\mathbf{m}_s \perp \mathbf{m}_u$, then the magnitude of
the local moment at each site goes as $m
=\sqrt{|\mathbf{m}_u|^2+|\mathbf{m}_s|^2}$, with $|\mathbf{m}_s| \sim (C_s
H\sin\alpha)^{\frac{1}{3}}$ and $|\mathbf{m}_u| \sim C_u H$. Here $C_s,C_u$ are some constants, and $\alpha$ is the
angle between $\mathbf{H}$ and $\mathbf{D}$. 
This finding implies the existence
of an angle dependent crossover field $H_c \sim \sqrt{C_s/C_u^3} \propto\sqrt{J|\mathbf{D}|}$ that would be on the
order of several Tesla for the 2D QSL ET salts at low temperature. For $H<H_c$ the local
magnetization is dominated by the staggered component and $m\propto
H^{\beta_L}$, $\beta_L\sim\frac{1}{3}$, while for $H>H_c$ the uniform component
dominates and $m\propto H^{\beta_H}$, $\beta_H \sim 1$. We argue that recent
powder $\mu$SR studies\cite{pratt2011magnetic} of
$\kappa$-(ET)$_2$Cu$_2$(CN)$_3$ under magnetic field detected exactly this behaviour; a fit of the
0.8 K $\mu$SR line width $B_e$ to the powder-averaged expression $B_e \propto
\int d\Omega \ m$ (Fig. \ref{fig-muon} (a)) yields $C_s/C_u=0.91$, $\beta_L=0.33$, and
$\beta_H=1.01$, in exact agreement with the expected (1D) exponents. These
exponents are not exotic; $\beta_L\sim 1/\delta$ where $\delta = 3$ is the
mean-field value, while constant $\chi_u$ is typical of both ordered and
disordered AFM systems. Previous NMR measurements at high field also suggested
$\beta_H \sim 1$, consistent with this picture~\cite{shimizu2006emergence},
while the value of $\beta_H \sim 0.83$ extracted from the $\mu$SR data in 
\cite{pratt2011magnetic} was measured within the
crossover region, and is therefore anomalously low. In
$\kappa$-(ET)$_2$Cu$_2$(CN)$_3$, the existence of this crossover requires only
that the low-field susceptibilities follow $\chi_s \gg \chi_u$, i.e. $\chi(\pi,\pi)\gg\chi(0,0)$, which
is expected since the QSL borders N\'eel
order\cite{kaneko2014gapless,PhysRevB.89.174415,motrunich2005variational,tocchio2009spin}.
Upon increasing temperature, $\chi_s (T)$ is expected to decrease much more rapidly than $\chi_u(T)$, so that the $\mathbf{h}_s$ dominated region forms a dome in the $H-T$ plane below some temperature $T^*\propto |\mathbf{D}|$ (Region I, Fig. \ref{fig-muon}), as observed experimentally. Thus, while $H_c$ has been identified as an exotic quantum critical
point\cite{pratt2011magnetic}, consideration of the staggered field
$\mathbf{h}_s$ due to the DM-interaction suggests a more conventional
interpretation. In fact, together with $\mathbf{D}$, an applied field
generally promotes spinon confinement through
$\mathbf{h}_s$~\cite{kenzelmann2004bound,schulz1996dynamics,zheludev2002interacting},
i.e. field-induced $(\pi,\pi)$ order with a small value of $\mathbf{m}_s\perp
\mathbf{H}$. Such induced order has been observed in
$\kappa$-(ET)$_2$Cu[N(CN)$_2$]Cl for $T > T_N$\cite{kagawa2008field}.
This effect may also explain the
field-induced broadening of the NMR lines, which was previously attributed to
short range order generated by impurities~\cite{shimizu2006emergence}. In
contrast, field-induced order may be long-range, and therefore provide well
defined antiferromagnetic resonance modes, making low $T$ ESR measurements of
great interest.\footnote{We note that it has been suggested that a
uniform component of $\mathbf{D}$ could result in a splitting of the ESR
line in QSL states due to broken chiral symmetry, and provide a means of
probing the spinon Fermi
surface\cite{gangadharaiah2008spin,tretiakov2013spin,povarov2011modes,glenn2012interplay}.
Unfortunately, crystal symmetry forbids this effect in
$\kappa$-(ET)$_2$Cu$_2$(CN)$_3$, where no uniform component of
$\mathbf{D}$ exists.}

Finally, we consider the residual effects not
included in the staggered-field picture above; for a nonzero component of
$\mathbf{D}$ along the $2_1$ axis direction, the anisotropic terms cannot
be completely gauged away. This leads to residual Ising-like terms associated with zero-field spin gaps on the order of 
$\Delta_s\sim\mathbf{\Gamma}\sim |\mathbf{D}|^2/J \sim 10- 50$ mK. For $\kappa$-(ET)$_2$Cu[N(CN)$_2$]Cl, for
example, the zero-field magnon gap can be estimated from $\Delta_s \sim \mu_B
H_{sf} \sim 50$ mK where $H_{sf} \sim 0.1$ T is the spin-flop
field~\cite{pinteric1999magnetic,miyagawa1995antiferromagnetic}. Spinon gaps of
smaller size have been suggested for the spin-liquid candidate
$\kappa$-(ET)$_2$Cu$_2$(CN)$_3$ from $\mu$SR measurement ($\Delta_s =$ 3.5
mK)~\cite{pratt2011magnetic}. While a complete description of SOC effects on the
QSL is beyond the scope of this work, we emphasize that any complete theory of
these very small spin gaps cannot ignore SOC effects. The large separation of
energy scales $J\gg T^* \gg \Delta_s$ is naturally explained by the
relative weakness of SOC without requiring another small emergent energy scale. These observations are generic to light-element magnets, and relevant also to e.g. the Cu-based Kagome antiferromagnet Herbertsmithite ZnCu$_3$(OH)$_6$Cl$_2$, which shows similar field-induced behaviour\cite{jeong2011field}.

We conclude by questioning the common assumption that SOC is sufficiently weak
in organics to be safely ignored by asking {\it weak compared to what?} We have
highlighted several cases where there are no relevant energy scales to compete
with SOC other than $k_B T$ or $g\mu_BH$, and therefore SOC cannot be ignored
{\it a priori}. We have presented a simple description of SOC effects in such
materials, and an intuitive picture of the orientation of the DM-vectors. In
the organics, SOC is most relevant for the $\kappa$- and $\alpha$-phase salts.
For $\kappa$-phase BETS salts, a large spin-orbit gap at the zone edge would
provide large magnetic breakdown fields observable in quantum oscillation
experiments. Similar spin-orbit gaps also render impossible the realization of
a true zero-gap Dirac state in $\alpha$-(ET)$_2$I$_3$ and
$\alpha$-(BETS)$_2$I$_3$. For the insulating state of $\kappa$-phase ET salts,
SOC manifests as anisotropic exchange. This results in spin-canting through the
DM-interaction, introduces zero-field magnon or spinon gaps, and provides a
promising explanation of field-induced effects in the QSL candidate
$\kappa$-(ET)$_2$Cu$_2$(CN)$_3$.

In closing, we would like to acknowledge useful discussions with M. Kartsovnik, M. Dressel, K. Kanoda and M. Lang. This work was supported by the NSERC of Canada and the Deutsche Forschungsgemeinschaft (DFG) through project SFB/TRR49.

\bibliography{SOC}
\clearpage
\section{Supplemental Material}

\begin{figure}[t]
\includegraphics[width=0.95\linewidth]{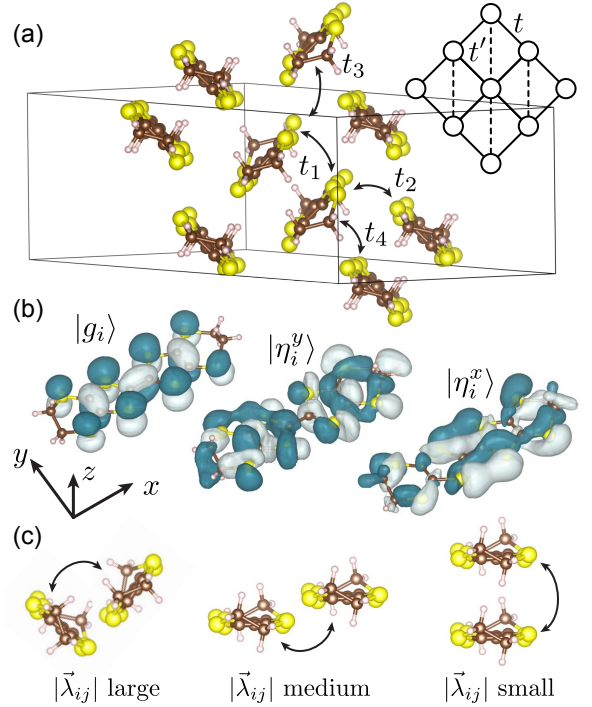}
\caption{\label{fig-orbs} (a) Definition of hopping integrals $t_{1-4}$ in the molecular basis for $\kappa$-phase salts, and corresponding integrals in the $t,t^\prime$ dimer basis. (b) ET HOMO $|g_i\rangle$ and pseudo-orbitals $|\eta_i^x\rangle$ and $|\eta_i^y\rangle$ with respect to the local coordinate system shown; $|\eta_i^z\rangle$ is omitted since $N_z \sim 0$. Only $|\eta_i^x\rangle$ has the correct symmetry to overlap with $|g_j\rangle$ on adjacent molecules, so that $\vec{\lambda}_{ij}$ is always nearly along the long-axis ($x$-axis) of the molecules. Optimal overlap of $|\eta_i^x\rangle$ and $|g_j\rangle$ is obtained for $90^\circ$ degree relative orientation of the molecules, as shown in (c), so that the magnitude $|\vec{\lambda}_{ij}|$ is maximized for this orientation.}
\end{figure}

{\it Calculation of hopping integrals:} As described below, in order to build
the SOC hopping integrals $\vec{\lambda}_{ij}$, it is necessary to first obtain
hopping integrals between all pairs of orbitals $\alpha$ and $\beta$ at
molecular sites $i$ and $j$ respectively (see Fig.~\ref{fig-orbs} (a)). These were approximated by
$t_{ij}^{\alpha\beta} = \langle \alpha_i|\mathcal{F}|\beta_j\rangle$, where
$\mathcal{F}$ is the single particle DFT operator computed using the local
quantum chemistry package ORCA at the B3LYP/def2-VDZ level via the following
procedure: (1) An initial calculation was performed for each molecular pair
$i,j$ in order to obtain  an initial set of Kohn-Sham orbitals
$|\Psi\rangle$ and eigenvalue matrix $\mathbf{E}$. These orbitals are
delocalized over both molecular sites, and serve as a poor basis for discussion
of intermolecular hopping. Therefore, (2) the obtained orbitals were projected
onto those of the isolated molecules, and reorthonormalized, to yield
$\mathbf{S}^{-1/2}|\Psi^\prime\rangle = \mathbf{P}|\Psi\rangle$, where
$\mathbf{P}$ is the projection matrix, and $\mathbf{S}$ is the overlap matrix.
Applying the same transformation to $\mathbf{E}$ yields the approximate
single-particle Hamiltonian in a site-local basis: $\mathbf{H}^\prime =
\mathbf{S}^{-1/2}\mathbf{P}\mathbf{E}\mathbf{P}\mathbf{S}^{-1/2}$. This
Hamiltonian has the unfortunate feature of including on-site terms
$t_{ii}^{\alpha\beta}$ due to crystal field effects. (3) In order to remove
these terms, $\mathbf{H}^\prime$ was selectively diagonalized at each site to
yield $\mathbf{H}^{\prime\prime} = \mathbf{U}^\dagger \mathbf{H}^\prime
\mathbf{U}$. The final hopping integrals are then given as the entries of
$\mathbf{H}^{\prime\prime}$. For the organics, this method has proved reliable,
and agrees very well with hopping integrals extracted from solid-state DFT
calculations in terms of a Wannier function
basis \cite{guterding2015influence,kandpal2009revision,PhysRevB.89.045102}.
 Using the obtained hopping integrals, the spin-orbit hopping elements
$\vec{\lambda}_{ij}$ were computed in the first order approximation:
\begin{align}
\vec{\lambda}_{ij} \sim \frac{1}{i}\sum_{n} \frac{\langle g_i|\vec{\mathcal{L}}_i|n_i\rangle}{\epsilon_g-\epsilon_n} t_{ij}^{ng} + t_{ij}^{gn} \frac{\langle n_j|\vec{\mathcal{L}}_j|g_j\rangle}{\epsilon_g-\epsilon_n} \label{eqn-7}
\end{align}
where $|g_i\rangle$ is the HOMO of an isolated ET molecule $i$; the sum over $n$ is carried out over all other valence orbitals on a given ET molecule, and $\mathcal{L}$ denotes the mean field SOC operator implemented in ORCA, and described in \cite{hess1996mean,neese2005efficient}. Computed values are shown in Table \ref{table-hops}; $\vec{\lambda}_1=\vec{\lambda}_3=0$ due to the presence of a crystallographic inversion centre.

{\it Orientation of $\vec{\lambda}_{ij}$:} 
For any pair of molecules, the orientation of $\vec{\lambda}_{ij}$ can be
understood by reinterpreting eq'n (\ref{eqn-7}) as defining a set of
pseudo-orbitals $|\eta_i^\mu\rangle =  \frac{1}{N_\mu}\sum_n |n_i\rangle
\frac{\langle n_i|\mathcal{L}_i^\mu|g_i\rangle}{\epsilon_g-\epsilon_n}$, where
$\mu=\{x,y,z\}$, and $N_\mu=  \sum_n  \frac{\langle
n_i|\mathcal{L}_i^\mu|g_i\rangle}{\epsilon_g-\epsilon_n }$ is a
direction-dependent normalization. A similar construction was discussed in 
\cite{thirunavukkuarasu2015pressure,winter2015magnetic}. Including SOC at
first order, the local spin-orbital doublet ground state of each molecule is
composed of:\begin{align}
|\uparrow_i\rangle \equiv & \  |g_i,\uparrow\rangle + N_z |\eta_i^z,\uparrow\rangle + N_x |\eta_i^x,\downarrow\rangle+iN_y|\eta_i^y,\downarrow\rangle \\
|\downarrow_i\rangle \equiv & \  |g_i,\downarrow\rangle - N_z |\eta_i^z,\downarrow\rangle + N_x |\eta_i^x,\uparrow\rangle-iN_y|\eta_i^y,\uparrow\rangle
\end{align}
For the organics, $N_\mu$ is typically very small, indicating only small
perturbations to the states. The small spin-orbit hopping elements
$\vec{\lambda}_{ij}$ represent the modification of the hopping via mixture of
the $|\eta_i^\mu\rangle$ pseudo-orbitals into the HOMO state. 
Thus, the components
$ \lambda_{ij}^\mu = \frac{N_\mu}{i} \left( t_{ij}^{\eta^\mu g} + t_{ij}^{g\eta^\mu}\right)$ are the weighted sum of hopping integrals between the HOMO and pseudo-orbitals $|\eta_i^\mu\rangle$ on adjacent sites.
In this way, studying the spatial distributions of $|\eta_i^\mu\rangle$ allows for understanding  the direction of $\vec{\lambda}_{ij}$\cite{thirunavukkuarasu2015pressure,winter2015magnetic}. Computed $|\eta_i^\mu\rangle$ pseudoorbitals for ET are shown in Fig. \ref{fig-orbs}(b). 

 \begin{table*}[t]
\caption {\label{table-hops} Hopping and SOC integrals computed for a variety of $\kappa$-phase salts based on crystal structures at the listed temperature $T$ (K). Crystal structures were obtained from the indicated references. All units are meV unless otherwise stated. For $Pnma$ materials, $\lambda_n$ are printed in $(a,b,c)$ coordinates, while for $P2_1/c$ materials, the coordinates are $(a,b,c^*)$. $\vec{\lambda}_1=\vec{\lambda}_3=0$ due to the presence of a crystallographic inversion centre. $\Delta_\text{eff}=2|\vec{\lambda}_\mathbf{q}|$, where $\mathbf{q}$ is $\mathbf{k}$-point where the Fermi surface meets the Brillouin zone boundary.}
\begin{ruledtabular}
\begin{tabular}{ccccccccccc}
Material&$T$(K)&Ref&Space Group&$t_1$&$t_2$&$t_3$&$t_4$&$\lambda_2$&$\lambda_4$&$\Delta_\text{eff}$\\
\hline
$\kappa$-(ET)$_2$Cu[N(CN)$_2$]Cl&100&\cite{hiramatsu2015quantum}&{\it Pnma}&187&100&59.0&33.9&(-0.03, +0.12, +0.10)&(-0.88, -0.99, -0.18)&1.6\\
$\kappa$-(ET)$_2$Cu$_2$(CN)$_3$&5&\cite{jeschke2012temperature}&{\it P$2_1/c$}&192&78.1&90.8&11.1&(+0.22, +0.11, 0.00)&(+0.90, +0.33, +0.32)&0.9\\
$\kappa$-(ET)$_2$Ag$_2$(CN)$_3$&300&\cite{blank2}&{\it P$2_1/c$}&163&72.8&66.9&20.0&(-0.15, -0.15, -0.15)&(-0.82, -0.24, -0.81)&1.2\\
$\kappa$-(ET)$_2$B(CN)$_4$&100&\cite{yoshida2015spin}&{\it Pnma}&195&36.3&118&25.5&(-0.14, -0.04, -0.18)&(+0.56, +1.81, +0.19)&1.2\\
$\kappa$-(BETS)$_2$Cu[N(CN)$_2$]Br&95&\cite{[Unpublished; available online at the Indiana University Molecular Structure Center: 
http://www.iumsc.indiana.edu] blank}&{\it Pnma}&229&126&51.4&35.8&(1.6, 2.5, -0.7)&(2.5, 2.4, 1.6)&4.7\\
$\kappa$-(BETS)$_2$GaCl$_4$&105&\cite{montgomery1996characterization}&{\it Pnma}&195&42.0&116&38.6&(-0.1, 0.9, 0.5)&(1.8, 7.8, 1.4)&5.5\\
\end{tabular}
\end{ruledtabular}
\end{table*}

For the purpose of discussion, we take the local $z$-axis to be perpendicular
to the molecular plane, while the $x$-axis is along the long axis of the
molecule, parallel to the central C-C bond. In these coordinates, the HOMO
$|g_i\rangle$ is composed of a linear combination of $p_z$ orbitals, so that
$\mathcal{L}_i^z|g_i\rangle \sim 0$. For this reason, $N_z\sim 0$, and one can
expect $\lambda_{ij}^z\sim 0$. In contrast, $|\eta_i^x\rangle$ and
$|\eta_i^y\rangle$ pseudo-orbitals are primarily composed of $p_x$ and $p_y$
orbitals; the effect of SOC is to mix the $\pi$- and $\sigma$- orbitals.
Comparing $|\eta_i^x\rangle$ and $|\eta_i^y\rangle$ reveals that only the
former has significant density extending beyond the edges of the ET molecules
and has the correct symmetry to overlap with the HOMO on adjacent molecules.
For this reason, $|t_{ij}^{\eta^x g}| > |t_{ij}^{\eta^y g}|$. From these
observations, we can conclude that $\vec{\lambda}_{ij}$ tends to point along
the long-axis of the ET molecules, regardless of the details of the crystal
packing. The magnitude of $\vec{\lambda}_{ij}$ is also enhanced for large
overlap between $p_z$ and $p_y$ orbitals on adjacent molecules, which occurs
for molecular planes oriented 90$^\circ$ from one another (see Fig.~\ref{fig-orbs} (c)). This observation
suggests the strongest SOC effects occur for $\alpha$-, $\theta$- and
$\kappa$-phase salts.

\begin{figure}[t]
\includegraphics[width=\linewidth]{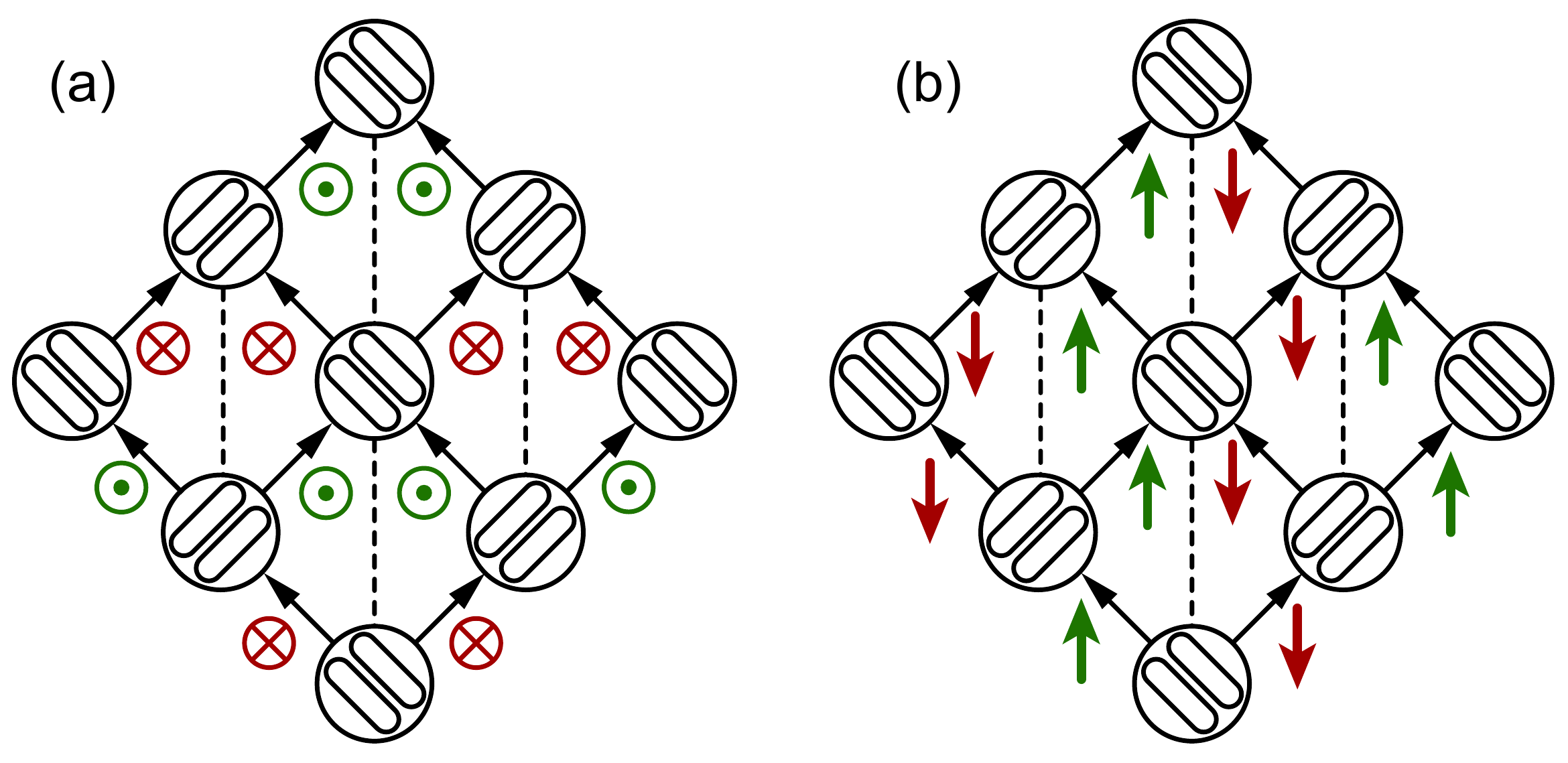}
\caption{\label{fig-DM} Pattern of $\mathbf{D}_{ij}$ vectors, showing the components perpendicular (a) and parallel (b) to the $2_1$ axis (i.e. the $ab$-plane, and $c$-axis, respectively, for $Pnma$ materials). Bonds are drawn as arrows with the head representing site $j$ and tail site $i$.}
\end{figure}

{\it Magnetic interactions:} In the $\kappa$-phase, in the large $U$ limit, a single hole is localized to each ET dimer, and occupies the anti-bonding combination of molecular HOMOs, given by $|a\rangle = (|g_1\rangle + |g_2\rangle)/\sqrt{2}$. Here, molecules 1 and 2 belong to the same dimer. The corresponding bonding combination $|b\rangle = (|g_1\rangle - |g_2\rangle)/\sqrt{2}$ is doubly occupied by electrons. In this basis, we consider a Hamiltonian that is the sum of, respectively, molecular orbital energies, on-site Coulomb repulsion, Hund's coupling, pair hopping, kinetic hopping, and nearest neighbour Coulomb repulsion:
\begin{align}
\mathcal{H} = \sum_i \mathcal{H}_{\epsilon} +\mathcal{H}_{U} +\mathcal{H}_{J} +\mathcal{H}_{\text{PH}} +\sum_{ij} \mathcal{H}_{\text{hop}}+\mathcal{H}_V \label{eq-Hubbard}
\end{align}
where $i$ labels a particular dimer site. Of these, the orbital energies in the hole picture are simply $\epsilon_{a} = -t_1, \epsilon_b = +t_1$, so that $\mathcal{H}_{\epsilon} = 2t_1 n_{i,b}$, where $t_1$ is the intradimer hopping integral (Fig.~\ref{fig-orbs} (a)). The interaction terms are given, in terms of hole operators, by:
\begin{align}
\mathcal{H}_{U} = & \ U n_{i,a}n_{i,b} +U \sum_\alpha n_{i,\alpha,\uparrow}n_{i,\alpha,\downarrow}  \\
\mathcal{H}_{J} = & \ J_H\sum_{\sigma,\sigma^\prime}c_{i,a,\sigma}^\dagger c_{i,b,\sigma^\prime}^\dagger c_{i,a,\sigma^\prime} c_{i,b,\sigma} \\
\mathcal{H}_{\text{PH}} =& \ J_H\sum_{\sigma,\sigma^\prime}c_{i,a,\sigma}^\dagger c_{i,a,\sigma^\prime}^\dagger c_{i,b,\sigma^\prime} c_{i,b,\sigma} \\
\mathcal{H}_V = & \ V\sum_{\alpha,\beta} n_{i,\alpha} n_{j,\beta}
\end{align}
where $n_{i,\alpha} = n_{i,\alpha,\uparrow}+n_{i,\alpha,\downarrow}$ is the number of holes at dimer site $i$, occupying the $\alpha$ orbital. We take $U = 0.55$ eV, $J_H = 0.2$ eV, and $V = 0.15$ eV. These values are consistent with those computed in \cite{nakamura2012ab}, but have been scaled by a factor $\sim 2/3$ in order to better match the experimental magnetic interactions. Hopping integrals in the $\{a,b\}$ basis can be obtained from the molecular HOMO-HOMO hopping integrals via: $t_{aa} =\pm t_{ab}= \frac{1}{2}(t_2+t_4)$, $t_{bb}= \pm t_{ba} =\frac{1}{2}(t_2-t_4)$, $t_{aa}^\prime = t_{ba}^\prime = \frac{1}{2}t_3$, and $t_{ab}^\prime =t_{bb}^\prime = -\frac{1}{2}t_3$. By convention, the $t$ bonds form a square lattice, while the $t^\prime$ bonds occupy the diagonals, as shown in Fig. \ref{fig-orbs}. 

Up to second order in interdimer hopping, but exact to all orders in $\{t_1,U,V,J_H\}$, the interactions are:
\begin{align}
\mathcal{H} = & \ \sum_{ij} J_{ij} \ \mathbf{S}_i \cdot \mathbf{S}_j + \mathbf{D}_{ij} \cdot \mathbf{S}_i \times \mathbf{S}_j + \mathbf{S}_i \cdot \mathbf{\Gamma}_{ij} \cdot \mathbf{S}_j
\end{align}
with:
\begin{align}
J_{ij} =& \ \frac{4(t_{ij}^{aa})^2}{U_\text{eff}}-\frac{2J_H[(t_{ij}^{ab})^2+(t_{ij}^{ba})^2]}{(2t_1+U-V)^2-J_H^2} \label{eqn-8}\\
\mathbf{D}_{ij} =& \  \frac{4t_{ij}^{aa}\vec{\lambda}_{ij}^{aa}}{U_\text{eff}}-\frac{2J_H(t_{ij}^{ab}\vec{\lambda}_{ij}^{ab}-t_{ij}^{ba}\vec{\lambda}_{ij}^{ba})}{(2t_1+U-V)^2-J_H^2} \\
\mathbf{\Gamma}_{ij} =& \   \frac{2\vec{\lambda}_{ij}^{aa}\otimes \vec{\lambda}_{ji}^{aa}}{U_\text{eff}}-\frac{J_H(\vec{\lambda}_{ij}^{ab}\otimes \vec{\lambda}_{ji}^{ba}+\vec{\lambda}_{ij}^{ba}\otimes \vec{\lambda}_{ji}^{ab})}{(2t_1+U-V)^2-J_H^2} 
\end{align}
where the effective Coulomb repulsion is:
\begin{align}
U_\text{eff} = & \ U-V-\frac{J_H^2}{4t_1+U-V} \sim 0.35\text{ eV}
\end{align}
\begin{table*}[t]
\caption {\label{table-Js_ED} Computed exchange interactions $J$ and DM vectors $\mathbf{D}$; all values are in units of K. Values for the $Pnma$ salts are in the coordinate system $(a,b,c)$, while the $P2_1/c$ values are with respect to $(a,b,c^*)$, comparison between perturbation theory results and cluster exact diagonalization on up to eight molecular sites.}
\begin{ruledtabular}
\begin{tabular}{c|cccc|cccc}
Material&\multicolumn{3}{c}{Perturbation theory}&\multicolumn{4}{c}{Exact Diagonalization}\\
		&$J$&$J^\prime$&$\mathbf{D}$&$\phi_a$&$J$&$J^\prime$&$\mathbf{D}$&$\phi_a$\\
\hline
$\kappa$-(ET)$_2$Cu[N(CN)$_2$]Cl
&523&96.2&(-3.51, -3.34, -0.26)&43.5
&482.0&164.7&(-3.65,	-3.58,	-0.17)&44.5\\
$\kappa$-(ET)$_2$Cu$_2$(CN)$_3$
&227&228&(+2.87, +1.13, +0.81)&21.5
&227.5&268.1&(+3.30,	+0.94,	+0.99)&13.8\\
$\kappa$-(ET)$_2$Ag$_2$(CN)$_3$
&249&122&(-2.58, -1.05, -2.58)&22.2
&250.2&157.8&(-2.93,	-0.92,	-2.93)&17.4\\
$\kappa$-(ET)$_2$B(CN)$_4$
&113&386&(+0.74, +3.20, +0.01)&77.0
&131.1&365.9&(+1.03,	+4.17,	-0.08)&76.2\\
\end{tabular}
\end{ruledtabular}
\end{table*}
\begin{table*}[t]
\caption {\label{table-Gs_PT_ED} Computed anisotropic exchange interactions $\Gamma$; all values are in units of mK. Values for the $Pnma$ salts are in the coordinate system $(a,b,c)$, while the $P2_1/c$ values are with respect to $(a,b,c^*)$, comparison between perturbation theory results and exact diagonalization on four-site clusters.}
\begin{ruledtabular}
\begin{tabular}{c|cc}
Material&Perturbation theory&Exact Diagonalization\\
		&$\Gamma$&$\Gamma$\\
\hline
$\kappa$-(ET)$_2$Cu[N(CN)$_2$]Cl
&$\begin{pmatrix}
5.14	&14.56	&1.44 \\
14.56	&4.48	&1.40 \\
1.44	&1.40	&-9.62
\end{pmatrix}$
&$\begin{pmatrix}
4.50	&11.84	&0.35 \\
11.84	&3.42	&0.29 \\
0.35	&0.29	&-7.91
\end{pmatrix}$
\\
$\kappa$-(ET)$_2$Cu$_2$(CN)$_3$
&$\begin{pmatrix}
12.56	&8.35	&6.34 \\
8.35	&-5.59	&2.47 \\
6.34	&2.47	&-6.97
\end{pmatrix}$
&$\begin{pmatrix}
13.32	&6.26	&6.44 \\
6.26	&-6.62	&1.84 \\
6.44	&1.84	&-6.70
\end{pmatrix}$ \\
$\kappa$-(ET)$_2$Ag$_2$(CN)$_3$
&$\begin{pmatrix}
4.70	&5.08	&15.59 \\
5.08	&-9.36	&5.08 \\
15.59	&5.08	&4.66
\end{pmatrix}$
&$\begin{pmatrix}
4.61	&5.08	&15.59 \\
5.08	&-9.23	&5.08 \\
15.59	&5.08	&4.62
\end{pmatrix}$ \\
$\kappa$-(ET)$_2$B(CN)$_4$
&$\begin{pmatrix}
-16.45	&13.51	&0.20	\\
13.51	&36.18	&0.85 \\
0.20	&0.85	&-19.70	
\end{pmatrix}$
&$\begin{pmatrix}
-18.53	&15.12	&-0.54	\\
15.12	&40.78	&-1.89 \\
-0.54	&-1.89	&-22.25	
\end{pmatrix}$ \\
\end{tabular}
\end{ruledtabular}
\end{table*}
The specific orientation of all $\mathbf{D}_{ij}$ are shown in Fig. \ref{fig-DM}. We note that it is conventional to estimate $J_{ij} \sim 4(t_{ij}^{aa})^2/(2t_1)$, which yields similar results because it happens accidentally that $U_\text{eff} \sim 2t_1$. Consideration of the above exact second order expressions demonstrates that this relationship does not hold generally\footnote{B. J. Powell, personal communication.}. The original motivation for this relationship is to assume a large bare Coulomb repulsion for a single ET molecule, while neglecting all other interactions. This is equivalent to the choice $U=J_H\gg t_1$, and $V = 0$. In this limit, the energy cost for transfer of a hole to an adjacent dimer is indeed $ \sim 2t_1$, but $J_{ij}$ is strongly suppressed due to the near degeneracy of singlet and triplet excited states with two holes on the same dimer. For this reason, the conventional estimate $J_{ij} \sim 4(t_{ij}^{aa})^2/(2t_1)$ is essentially wrong, but accidentally yields reliable values.

Given that the studied materials are in close proximity to the Mott transition, it has been suggested that higher order corrections also play a significant role in the magnetic response.
%
In order to incorporate higher order terms, we therefore computed the interactions via linked cluster expansion up to four dimer sites (eight molecular sites). 
In this approach, we interpret the exchange interactions as observables, which may be computed by summing over clusters $\{ C \}$:
\begin{align}
J = \sum_{\{C\}} \tilde{J}_C
\end{align}
where $\tilde{J}_C$ is the reduced contribution from cluster $C$, obtained by subtracting all subcluster contributions:
\begin{align}
\tilde{J}_C = J_C - \sum_{\{C^\prime\}\in C}\tilde{J}_{C^\prime}
\end{align}
where $J_C$ is the actual value of the exchange constant measured on cluster $C$, and the summation is carried over all subclusters of $C$. From the perspective of perturbation theory, each $\tilde{J}_C$ contains the terms that include hopping of particles between every site in $C$ (at all orders in $t/U$). The series is convergent, however, because the lowest order terms in each $\tilde{J}_C$ go as $U(t/U)^N$, where $N$ is the number of sites in cluster $C$. For each cluster, $J_C$ is measured via exact diagonalization of the Hubbard model~\eqref{eq-Hubbard} as in the previous works \cite{winter2016challenges, riedl2016}. As expected, the low energy states consist of one hole per dimer, localized in the antibonding state with only small contributions from other states. Therefore, all contributions including double occupancy are projected out, and the low-energy states are reorthogonalized to yield a mapping between the exact low-energy states of the cluster, and the desired pure spin states. The Hamiltonian that results from this procedure is the desired spin Hamiltonian, from which the exchange parameters can be directly read. The resulting spin Hamiltonian preserves all symmetries of the cluster $C$, reproduces the low-energy spectrum of the Hubbard model on the cluster, and converges exactly to the perturbation theory result in the $t/U\rightarrow 0$ limit. This method is similar to the so-called CORE (contractor renormalization) technique, where the effective Hamiltonian results from calculating effective models by truncating local degrees of freedom, and the final effective Hamiltonian is the sum of connected clusters, with contributions from embedded sub-clusters substracted \cite{morningstar1994core, morningstar1996core, Yang2012core}.



\begin{figure}[t]
\includegraphics[width=\linewidth]{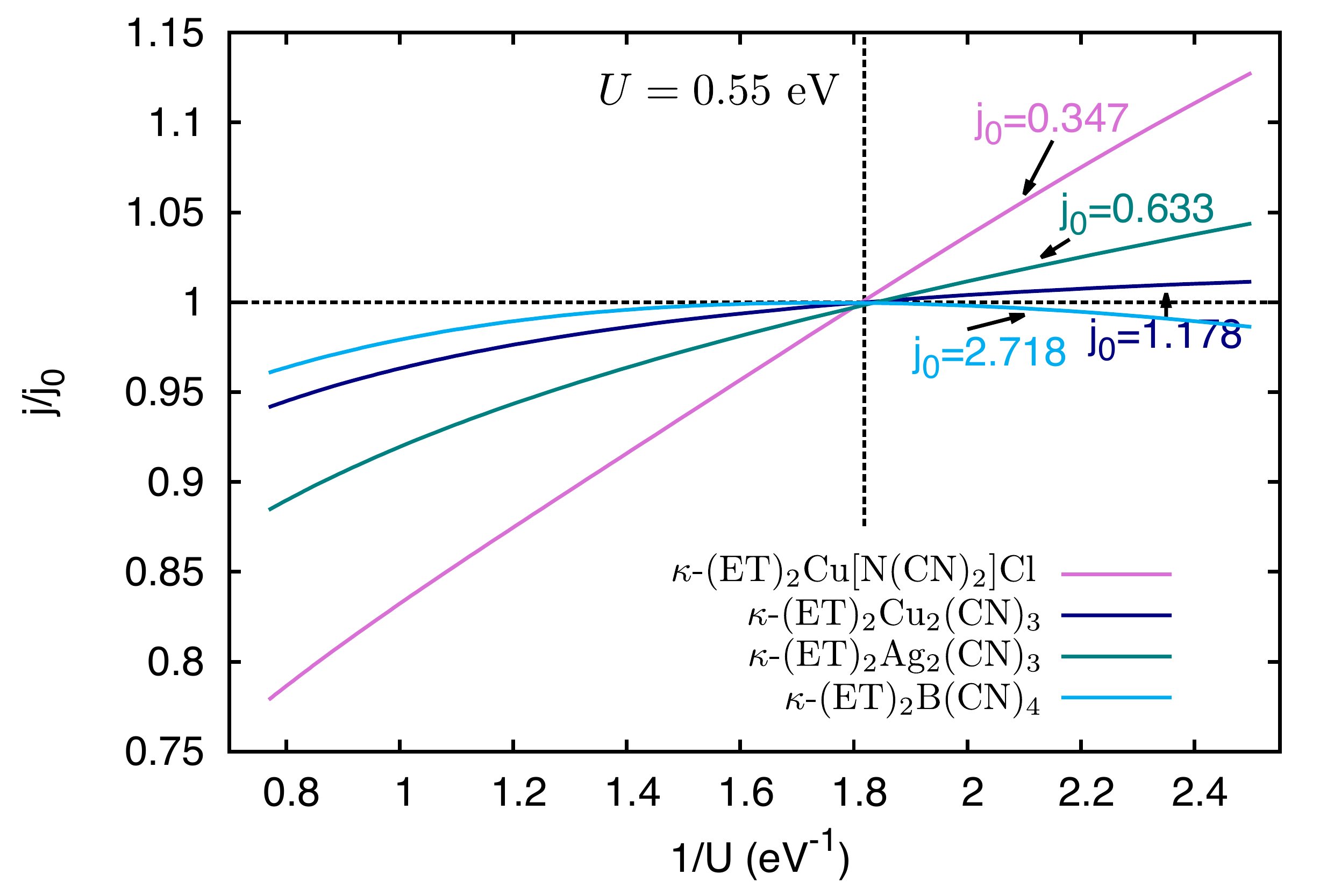}
\caption{\label{fig-J-JP} Ratio of the Heisenberg exchange parameters $j=J^\prime/J$, rescaled with respect to the value $j_0$ obtained with $U=0.55$ eV (see Tab.~\ref{table-Js_ED}), for varied Hubbard $U$ parameters. The exchange parameters are determined with exact cluster diagonalization up to eight molecules and we used the model parameters $V=0.25 \ U$ and $J_H=0.4375 \ U$ following the ratios used in Ref.~\cite{nakamura2012ab}.}
\end{figure}

For $\kappa$-(ET)$_2$Cu[N(CN)$_2$]Cl, the $J^\prime$ determined from perturbation theory is given by $96.2$ K. The cluster expansion value, taking only clusters up to two dimer sites is $J^\prime=94.6$ K, which agrees well with the perturbation theory value. Taking all clusters up to four dimer sites increases the estimate to $J^\prime = 164.7$ K, suggesting significant contributions from higher order corrections, which increases the $J^\prime/J$ ratio. On the other hand in the quasi 1D material $\kappa$-(ET)$_2$B(CN)$_4$ higher order corrections lead to a reduction of this ratio. In Fig.~\ref{fig-J-JP} we illustrate the ratio $j=J^\prime/J$ in terms of the values $j_0$ as a function of $1/U$. Close to the Mott transition, in the regime of weak correlations, where higher order terms become more important, $J$ and $J^\prime$ approach towards each other and the frustration is increased for all investigated materials, except for the spin-liquid candidate $\kappa$-(ET)$_2$Cu$_2$(CN)$_3$.


{\it Scaling of $H_c$ and $T^*$:} In the low temperature limit, we assume that $\chi_s \sim |\mathbf{h}_s|^{-2/3}$, and $\chi_u$ is constant. Thus, following the argumentation for 1D chains in \cite{affleck1999field}, the staggered an uniform magnetizations follow:\\
\begin{align}
m_u \propto \frac{|\mathbf{h}_u|}{J}\propto \frac{H}{J} \\
m_s \propto \left(\frac{|\mathbf{h}_s|}{J}\right)^{1/3} \propto \left(\frac{|\mathbf{D}|}{J^2}\right)^{1/3}H^{1/3}
\end{align}
The crossover field $H_c$, where $m_u \approx m_s$ can thus be obtained:
\begin{align}
H_c \propto \sqrt{|\mathbf{D}|J}
\end{align}
In the high temperature limit, $T \gg |\mathbf{h}_s|$, we expect the staggered susceptibility to follow a Curie Law, $\chi_s \propto T^{-1}$, while the uniform susceptibility may remain roughly constant. Thus:
\begin{align}
m_u \propto \frac{|\mathbf{h}_u|}{J}\propto \frac{H}{J} \\
m_s \propto \frac{|\mathbf{h}_s|}{T} \propto \frac{|\mathbf{D}|H}{JT}
\end{align}
The crossover temperature $T^*$, where $m_u \approx m_s$ can thus be obtained:
\begin{align}
T^* \propto |\mathbf{D}|
\end{align}
Thus, $H_c$ and $T^*$ limit the staggered field dominated region to relatively low field and temperatures. For temperatures close to $T^*$, the system should display $H/T$ scaling, implying a linear boundary between the spin-liquid and field-induced phases, as observed in $\kappa$-(ET)$_2$Cu[N(CN)$_2$]Cl \cite{pratt2011magnetic}. For temperatures $T\ll T^*$, we expect instead $T/H^{2/3}$ scaling, which would fit the results for the Kagome antiferromagnet Herbertsmithite ZnCu$_3$(OH)$_6$Cl$_2$ \cite{jeong2011field}. These results assume similar scaling forms as in the 1D antiferromagnetic chains, and could be refined by explicit estimation for 2D systems.

%

\end{document}